# Generating high-quality 3DMPCs by adaptive data acquisition and NeREF-based radiometric calibration with UGV plant phenotyping system

Pengyao Xie[1,2], Zhihong Ma[1,2], Ruiming Du[1,2], Xin Yang[1,2], Haiyan Cen [1,2]*

[1] College of Biosystems Engineering and Food Science, Zhejiang University, Hangzhou 310058, China

[2] Key Laboratory of Spectroscopy Sensing, Ministry of Agriculture and Rural Affairs, Hangzhou 310058, China

* Correspondence: hycen@zju.edu.cn; Tel: +86 571 88982527.

Highlights

1. An efficient method for next-best-view (NBV) estimation with high accuracy in the case of the limited camera field of view (FOV) was proposed.
2. A path planning algorithm based on hybrid particle swarm optimization (HPSO) and rapidly-exploring random trees (RRT) was proposed.
3. A radiometric calibration method under the natural light condition based on the neural reference field (NeREF) was proposed.
4. The accuracy of chlorophyll content estimation for plants such as perilla, tomato and rapeseed can be improved by using high-quality 3D multispectral point clouds (3DMPCs).

**Abstract:**

Fusion of three-dimensional (3D) and multispectral (MS) imaging data has a great potential for high-throughput plant phenotyping of structural and biochemical as well as physiological traits simultaneously, which is important for decision support in agriculture and for crop breeders in selecting the best genotypes. However, lacking of 3D data integrity of various plant canopy structures and low-quality of MS images caused by the complex illumination effects make a great challenge, especially at the proximal imaging scale. Therefore, this study proposed a novel approach for adaptive data acquisition and radiometric calibration to generate high-quality 3D multispectral point clouds (3DMPCs) of plants. An efficient next-best-view (NBV) planning method based on an unmanned ground vehicle (UGV) plant phenotyping system with a multi-sensor-equipped robotic arm was proposed to achieve adaptive data acquisition. The neural reference field (NeREF) was employed to predict the digital number (DN) values of the hemispherical reference for radiometric calibration. For NBV planning, the average total time for single plant at a joint speed of 1.55 rad/s was about 62.8 s, with an average reduction of 18.0% compared to the unplanned. The integrity of the whole-plant data was improved by an average of 23.6% compared to the fixed viewpoints alone. Compared with the ASD measurements, the average root mean square error (RMSE) of the reflectance spectra obtained from 3DMPCs at different regions of interest was 0.08 with an average decrease of 58.93% compared to the results obtained from the single-frame of MS images without 3D radiometric calibration. The 3D-calibrated plant 3DMPCs improved the predictive accuracy of partial least squares regression (PLSR) for chlorophyll content, with an average increase of 0.07 in the coefficient of determination ($R^2$) and an average decrease of 21.25% in RMSE. Our approach introduced a fresh perspective on generating high-quality 3DMPCs of plants under the natural light condition, enabling more precise analysis of plant morphological and physiological parameters.

## 1. Introduction

High-throughput plant phenotyping provides an unprecedented way to systematically evaluate plant development and functionality with the precise quantification of morphological, physiological, biochemical, and performance traits over the whole growth period. It can help on decision support in agriculture, for ecological diversity studies, and for crop breeding in the selection of superior genotypes to improve crop performance, and thus revolutionize the agriculture and breeding strategies to meet the future need of agricultural sustainable development (Freschet et al. 2018; Hu and Schmidhalter 2023). In the last years, proximal unmanned ground vehicles (UGVs) or robots are increasingly adopted as plant phenotyping platforms carried out with various types of imaging sensors (Jin et al. 2021; Xu and Li 2022). By designing different chassis configurations, UGVs can be easily adopted in indoor (Bao et al. 2019), greenhouse (Martin et al. 2021), and open-field environments (Perez-Ruiz et al. 2020). Compared to the unmanned aerial vehicles (UAVs), gantry systems, and convey belt systems, UGVs are more flexible to adjust the position and orientation to acquire high temporal, spatial, and spectral resolution data from organ- to plant-levels (Atefi et al. 2021). Additionally, sensors, robotic arms, and operation grippers can be easily integrated into the UGVs, making them more extensible for high-throughput plant phenotyping.

UGVs combined with multiple imaging sensors to collect comprehensive information related to plant phenotypic traits has progressed significantly with the fast development of low-cost, light-weight sensors, leading to an urgent need to advance data fusion technologies to take the full benefit of multi-sensor inputs (Brell et al. 2019; Sun et al. 2022). It is hypothesized that such platforms are ideal for generating three-dimensional multispectral point clouds (3DMPCs) for mapping spectral information at a 3D coordinate of discrete points, which are highly needed to reveal the physio-biochemical traits within the plant 3D structural domain (Behmann et al. 2016). However, due to the complexity of plant canopy structures and their interaction with light, acquisition of integral 3D data of various plant canopies and high-quality, well-calibrated spectral images becomes a main challenge at proximal phenotyping (Behmann et al. 2015). Acquiring multi-source image data by manually presetting fixed viewpoints (FVPs) or paths not only are relatively subjective, but it is also time-consuming as the whole process is semi-automatic (Liang et al. 2013). In addition, the plant occlusion, limited depth of field, and improper camera field of view (FOV) would add more uncertainties in the FVPs image acquisition, resulting in a low signal-to-noise ratio (SNR) in image data.

Adaptive data acquisition based on the active vision technology is an objective-based, automatic imaging process by perception of the operating environment through planned sensing (Chen et al. 2011). While the complex and variable plant structures, sensing uncertainties, and non-structural environments when dealing with plant phenotyping tasks place higher demands on the performance of adaptive algorithms (Araus et al. 2018). Suitable unobstructed viewpoints must be found for imaging sensors to directly obtain complete and accurate 3D data (Martin et al. 2021; Schor et al. 2017). The next-best-view (NBV) method iteratively estimates the next-best viewpoint based on the existing inferior data to provide the destination for path planning, and would be a possible solution for adaptive data acquisition in plant phenotyping tasks (Vasquez-Gomez et al. 2014). Among all the candidate viewpoints, the next-best viewpoint is able to balance the information gain against the movement cost for adaptive data acquisition, which requires the design of objective functions to represent the specific mathematical relationships. There are mainly 2D-based approaches and 3D-based approaches for information gain calculation. 2D-based approaches such as 3D move to see (3DMTS) (Lehnert et al. 2019) and Deep-3DMTS (Zapotezny-Anderson and Lehnert 2019) could increase the target size by estimating a gradient that represents the direction of the NBV. However, it requires processing a large amount of image data simultaneously with the high time cost and low visual servo rate. The 3D-based approaches use ray tracing to calculate the information gain for NBV planning (Foix et al. 2018). The complete pixel's frustum is usually adopted for accurate computation of ray tracing, which results in a slower speed. In addition, learning-based NBV estimation strategies require a large amount of sampling in advance to obtain enough training data (Wu et al. 2019). This process is time-consuming, poorly generalized, and not conducive to small samples. Normal view, generally considered to be an empirical result of view estimation with the maximum information gain, is quite easy to estimate. While the data obtained directly from the normal view is incomplete due to occlusion (Alenya et al. 2013; Alenya et al. 2011). Existing improvement used ray tracing to discriminate the occlusion of the best viewpoint and implemented incremental examination of the camera view sphere to explore other viewpoints around the normal (Gibbs et al. 2020). However, it was still computationally expensive. In summary, a fast estimation method for multiple unobstructed viewpoints based on the normal viewpoint correction is still lacking. Inspired by the path optimization method proposed by Tong et al. (2022), it is worth to investigate the heuristic sorting and path planning for the same batch of viewpoints for adaptive data acquisition.

Furthermore, the conversion of multispectral imaging signal to the reflectance by radiometric calibration is usually performed to obtain the reflectance spectral information that is related to the physiological and biochemical information of the plant. The quality of multispectral images can be affected by the illumination condition, plant structure, and observation angle, leading to the low SNR of spectral information. Radiometric calibration is required to eliminate such factors to obtain accurate multispectral point clouds in the 3D domain. A reference flat with uniform reflectance at different wavelengths is usually used for radiometric calibration, which is ideally suitable for a flat object. While the complex plant canopy structure makes it impossible to match the geometry of the plant's light field with that of the reference flat. Moreover, it is impractical to change the position of the reference flat to simulate the light interaction with plants. Physical modeling was proposed such as PROCOSINE, combining PROSPECT and ClOse-range Spectral ImagiNg of lEaves (CONSINE) to describe the leaf specular reflectance and local leaf orientation at the leaf scale (Jay et al. 2016; Proctor et al. 2021; Shi and Xiao 2021). The requirement of measuring different parameters for physical models make it difficult for practical applications. In addition, Zhang et al. (2020) developed a stereo reference-based radiometric calibration method that can only be used for a single leaf. Alternatively, we designed a radiometric calibration method for complex individual plants based on the hemispherical references under the controlled light condition (Xie et al. 2023). However, it is impossible to apply the previously proposed point-source light field representation directly to the outdoor scenes due to the multi-view image acquisition with different light conditions. Acquisition of multi-view images by changing camera poses under natural light conditions requires a new parallel light field representation, which are beyond the capabilities of existing approaches (e.g., neural 4D light field (Li et al. 2021), 6D Plücker light field (Sitzmann et al. 2021), handcrafted priors (Levin et al. 2010; Shi et al. 2014; Vagharshakyan et al. 2018) and deep learning-based light fields (Bemana et al. 2020; Kalantari et al. 2016; Mildenhall et al. 2019)). We proposed to use a novel variant of neural radiance field (NeRF) with the inputs of the 3D parallel light field features to implement a surface-based implicit neural representation of the hemispherical reference (Mildenhall et al. 2022; Tewari et al. 2020; Xie et al. 2022). To best of our knowledge, this is the first study that NeRF-like model was employed for accurately estimate the complete multi-view digital number (DN) values of the hemispherical reference with different light field features, thereby greatly improving the accuracy of stereo reference-based radiometric calibration.

Hence, this study aims to develop a new method to generate high-quality 3DMPCs by adaptive multi-sensor data acquisition and 3D radiometric calibration. The specific objectives of this study are: (1) to achieve NBV planning based on the proposed adaptive algorithm and the collision-free path planning algorithm, (2) to calibrate reflectance spectra under the natural light condition based on NeRF variant, and (3) to assess the performance of the proposed approach for chlorophyll content estimation using 3DMPCs of plants.

## 2. Materials and Methods

Figure 1 illustrates the whole process of generating the high-quality 3DMPC of plants. The process mainly consists of three steps, including image data acquisition, radiometric calibration, and data fusion. The RGB-D images of the plant and the solid Teflon hemispherical reference with the diameter of 100 mm were first acquired in the starting view to perform the NBV estimation for path planning. More RGB-D images and multispectral (MS) images from different viewpoints were then adaptively acquired before they were registered to generate the aligned image pairs. The 3D light field features and the DN values were extracted from the aligned image pairs next to prepare the dataset for training the Neural REference Field (NeREF). Finally, the calibrated reflectance images and the RGB-D images from multi-views were fused to construct the 3DMPC. Detailed descriptions of these three steps were provided in the following sub-sections.

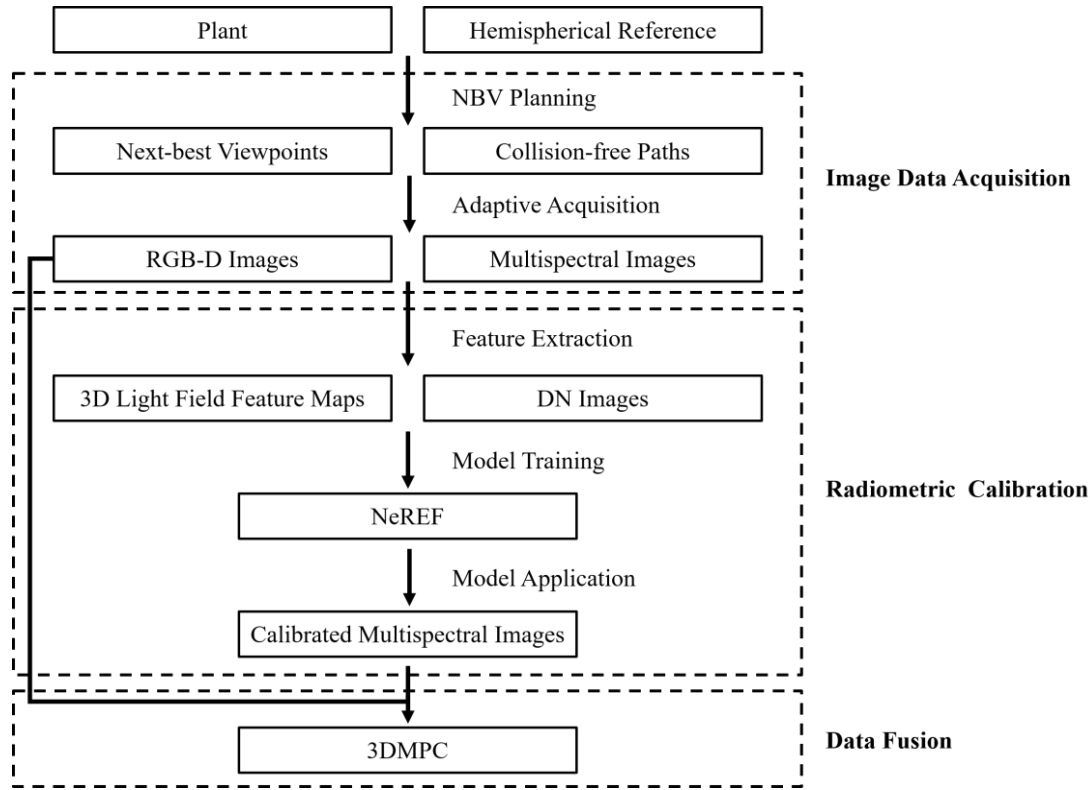

Figure 1. Flowchart of generating high-quality three-dimensional multispectral point cloud (3DMPC) of plants.

## 2.1. Data Acquisition

### 2.1.1. UGV Plant Phenotyping System

The schematic diagram of the UGV plant phenotyping system is shown in Figure 2. The system consisted of an UGV, a multi-sensor-equipped robotic arm (UR5, Universal Robots, Odense, Denmark), and a laptop to control the whole system. The sensors were mounted on the end-effector which included an RGB-D camera (Azure Kinect DK, Microsoft, Redmond, WA, USA), a MS camera (MQ022MG-CM, XIMEA, Munster, Germany), and a laser scanner (PRINCE775, SCANTECH, Hangzhou, China). The RGB-D camera worked with a positive hexagonal FOV of **75° × 65°** and a working range from 0.50 m to 3.86 m in "NFOV UNBINNED" mode to capture images with the size of 720 × 1280 pixels. The MS camera worked with a smaller rectangular FOV of **16°** to capture images with 25 bands at 650-950 nm and the image size of 216 × 409 pixels.

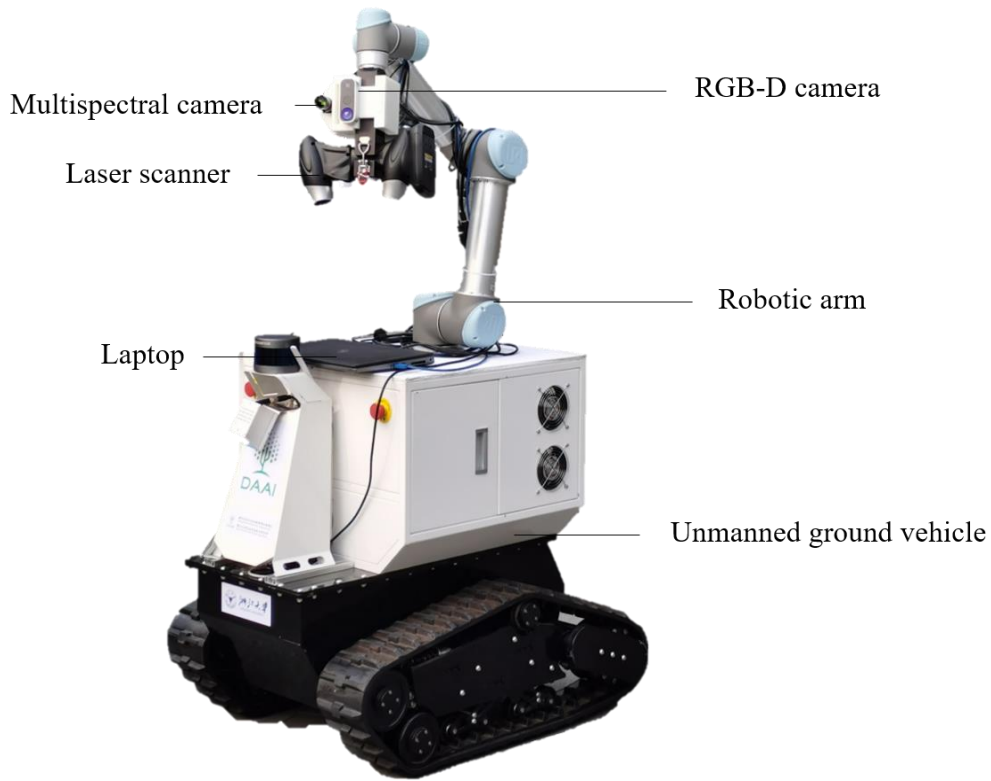

Figure 2. Schematic diagram of the mobile platform based on the multi-sensor-equipped robotic arm.

### 2.1.2. NBV Estimation

The schematic diagram of the adaptive data acquisition is presented in Figure 3. Firstly, the FVPs of the faces of the axis-aligned bounding box (AABB) were estimated based on a long-distance starting view in Step 1. By fusing the point clouds from each viewpoint, the plant contour with the partial details can be obtained. The vertical viewpoints (VVPs) and adaptive viewpoints (AVPs) were then estimated based on the fused whole-plant and leaf point clouds in Step 2. Leaf clustering was performed by using the region growing algorithm and principal component analysis (PCA) was applied for estimating the oriented bounding box (OBB) of the clustered point clouds. The VVPs were estimated based on the front of OBB (positive Z-axis direction). We chose the region that includes the most unobstructed viewpoints based on the morphology and topological relationships between the whole and parts of the plant (Algorithm 1 Line 1-5). The general contour of the plant can be described with the plant's convex hull whose larger polygonal surface may contain unobstructed viewpoints. The unit directional hemisphere represents all viewpoints in the hemispherical space where the main direction of the target was located. By calculating the difference between the projections of all polygonal surfaces and the rest of the plant on the hemisphere, all unobstructed viewpoints can be obtained. We selected the remaining projection containing the most unobstructed viewpoints as the search region for NBV and searched for viewpoints with the maximum canopy gap in this region (Algorithm 1 Line 6-14). The viewpoints with the maximum canopy gap cannot be calculated directly by using the geometric methods as the shape of the remaining projection region is usually irregular. Therefore, we sampled the irregular shape as the discrete point clouds with the fixed spacing and used a k-dimensional tree (kd-tree) to iteratively calculate the number of neighboring points in the incremental neighborhood of each point. Finally, the AVP with the most consecutive neighborhood points was obtained.

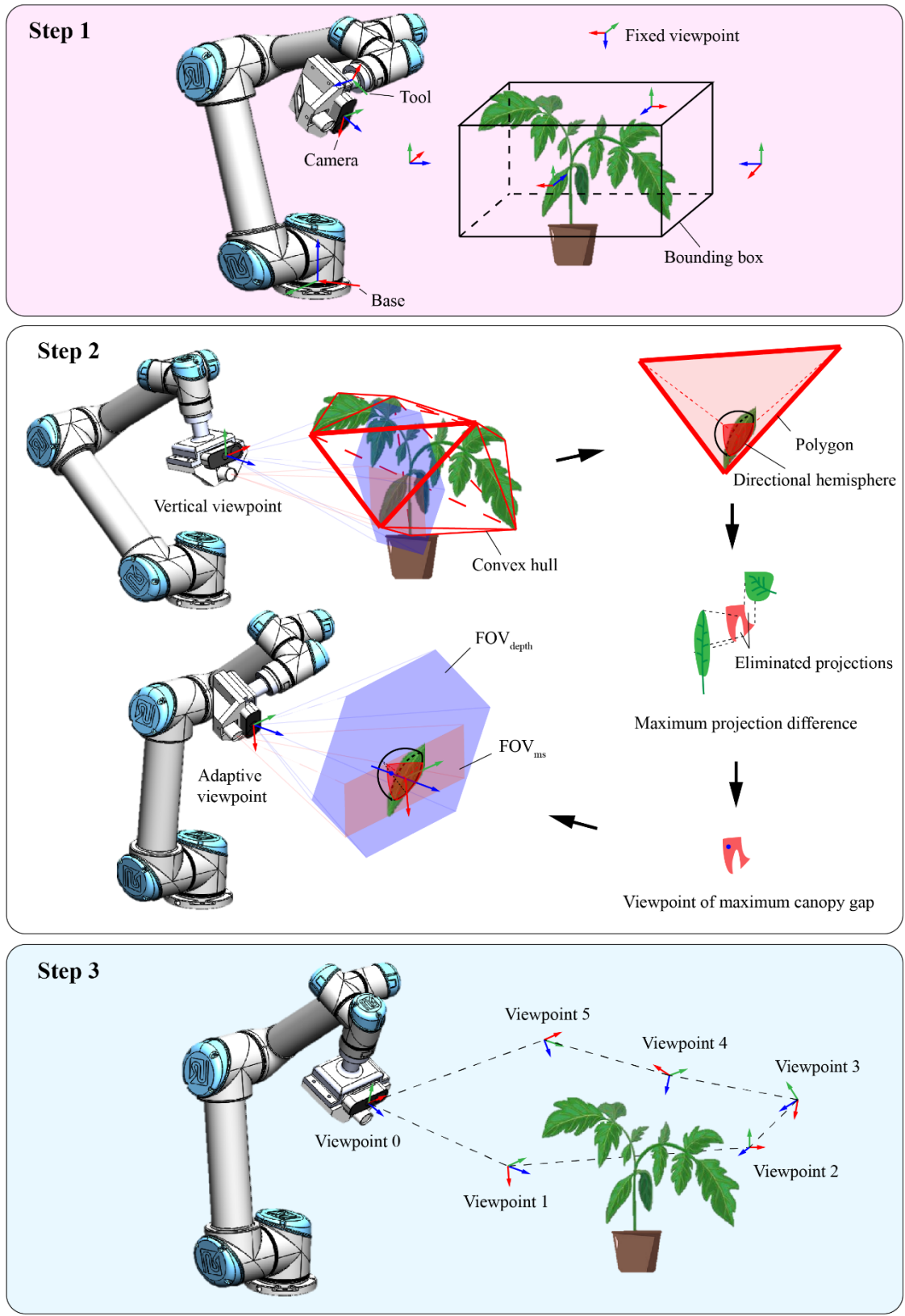

Figure 3. Schematic diagram of the adaptive data acquisition. The first two steps are next-best-view (NBV) estimation, and the third step is path planning.

**Algorithm 1** AVP estimation

**Input**:

Plant point cloud $P$ and its parts $\{P_i\}(i = 1, ..., M)$.

**Output**:

AVPs $\{v_i\}(i = 1, ..., M)$ for each part.

1. Calculate all polygons $\{T_j\}(j = 1, ..., K)$ of the plant convex hull and all unit directional hemispheres $\{H_i\}(i = 1, ..., M)$ of each part;
2. **For** $H_i \in H$ **do**
3.     **For** $T_j \in T$ **do**
4.         Calculate the projection difference on the hemisphere $D_j \leftarrow T_j^i - (P - P_i)^i$;
5.     **End**
6.     Downsample $D_{max}$ into the point set $\{d_k\}(k = 1, ..., Q)$ by spacing $\delta$;
7.     Initial the radius $r \leftarrow \delta + \Delta\delta$, $\Delta\delta \in (0, \delta)$ and the point set $b \leftarrow d$;
8.     **Repeat**
9.         **For** $b_q \in b$ **do**
10.             Search for points within radius $r$ in $d$ using the kd-tree and record the number of points $N_q$;
11.         **End**
12.         Update the radius $r \leftarrow r + \delta$ and the point set $b \leftarrow \{b_q | N_q = N_{max}\}$;
13.     **Until** $N_{max} = Q$ or $q_{max} = 1$;
14.     Randomly select a point in $b$ as the viewpoint $v_i$;
15. **End**
16. **Return** a set of AVPs $\{v_i\}(i = 1, ..., M)$;

We determined the orientation and the sight distance of the viewpoint as follows. The z-axis direction of the viewpoint was determined by the line between the center of the hemisphere and this point. The x-axis direction of the viewpoint was determined by a radius in the hemisphere plane, whose longitude was the same as the viewpoint. At last, the orientation calculation was completed by determining the y-axis direction according to the right-hand rule. We set the sight distance of all the viewpoints to 0.5 m according to the camera working range and FOV.

### 2.1.3. Path Planning

The paths among the sorted viewpoints were planned based on the hybrid particle swarm optimization (HPSO) and the rapidly-exploring random trees (RRT) in Step 3 as shown in Figure 3. The sorting among the viewpoints is considered as a traveling salesman problem (TSP) in the discrete domain and is suitable for optimization using heuristic algorithms. We incorporated the crossover and mutation of the genetic algorithm (GA) into the original PSO algorithm so that the current solution was crossed with the individual extreme value and the global extreme value respectively with a certain mutation rate to generate a new solution. After these processes, we adopted the probabilistic acceptance criteria in simulated annealing (SA) algorithm to allow a limited range of badness. We converted the Cartesian coordinates of the robotic arm at each viewpoint into the joint space coordinates and calculated the distance as the fitness of the HPSO algorithm:

$$Fitness = \sum_{n=1}^{N-1} \sqrt{\sum_{i=1}^{6}(J_i^n - J_i^{n+1})^2} \qquad (1)$$

where $J_i^n \in [0, 2\pi)$ is the coordinate of joint $i$ in the joint space of the robotic arm corresponding to viewpoint $n$, and $N$ is the number of viewpoints in a path. The proper number of iterations was then chosen to ensure its convergence. At last, the collision-free paths among the sorted viewpoints were planned using RRT proposed by LaValle and Kuffner (2000). RRT guides the tree to grow to the target by randomly selecting sampling points in the state space, and avoids the modeling of space by collision detection of sampling points. We firstly converted the scene point cloud into an octree-based grid map (Octomap). Then we utilized the RRT algorithm to generate waypoints and quickly find the initial path. The path was continuously optimized as the number of sampling points increased until the goal point was found or the number of iterations was reached.

### 2.2. Radiometric Calibration

We proposed a novel method named NeREF to implement a surface-based implicit representation of the hemispherical reference. NeREF encodes the full light field in the weights of a multi-layer perceptron (MLP) that maps an oriented ray to the DN value observed by that ray. Each ray is represented by a specific pixel-wise 3D light field feature. This frees us from the complex computation of the ray-surface intersection, and the volumetric & ray-marching based rendering. We also modified the light field representation to adapt to the natural light condition in this approach. For the point-source light field in the controlled light condition mentioned in our previous study (Xie et al. 2023), the 3D light field features can be represented as $(I_1, I_2, I_3, V_1, V_2, V_3, n_1, n_2, n_3)$, and the constraint is $\|(n_1, n_2, n_3)\| = 1$. Here, $(I_1, I_2, I_3)$, $(V_1, V_2, V_3)$, and $(n_1, n_2, n_3)$ are the direction vector of the incident light, the observation direction vector, and the normal vector of the surface, respectively. In the natural light condition, the light can be regarded as the combination of parallel light $I$ and diffuse light $I_0$. The incidence and observation distance factors can be neglected based on the assumption that parallel light does not decay with the distance and that the medium for light propagation is perfectly transmissive. Besides, since the incident angle of the parallel light is constant and that of diffuse light from the environment is omnidirectional, both can be described as distribution functions related to the observation only, without the need to consider incidence-related factors. Thus, the 3D light field features can be represented as $(v_1, v_2, v_3, n_1, n_2, n_3)$, and the constraint is $\|(v_1, v_2, v_3)\| = \|(n_1, n_2, n_3)\| = 1$. Here, $(v_1, v_2, v_3)$ and $(n_1, n_2, n_3)$ are the observation direction vector and the normal vector of the surface, respectively. Figure 4a presents the simulation of plant leaves and hemisphere tangent to each other and the relevant 3D light field features in the natural light condition. In this case, the plant and the reference at the tangent point share the same 3D light field features, so that the DN value of the reference at the tangent point can be directly used for the plant radiometric calibration. Figure 4b shows the different stages of NeREF modeling and details of the reflectance calculation method. We selected the 3D light field features of the reference under the natural light condition as the input and selected the 25-band spectral DN value as the output. The trained NeREF was then applied to predict the DN value with the 3D light field features of the plant as input. Finally, the calibrated MS reflectance image of the plant was obtained by calculating the percentage ratio of plant DN and reference DN pixel by pixel. It is worth mentioning that the effect of dark current on all DN values has been eliminated before the above processing.

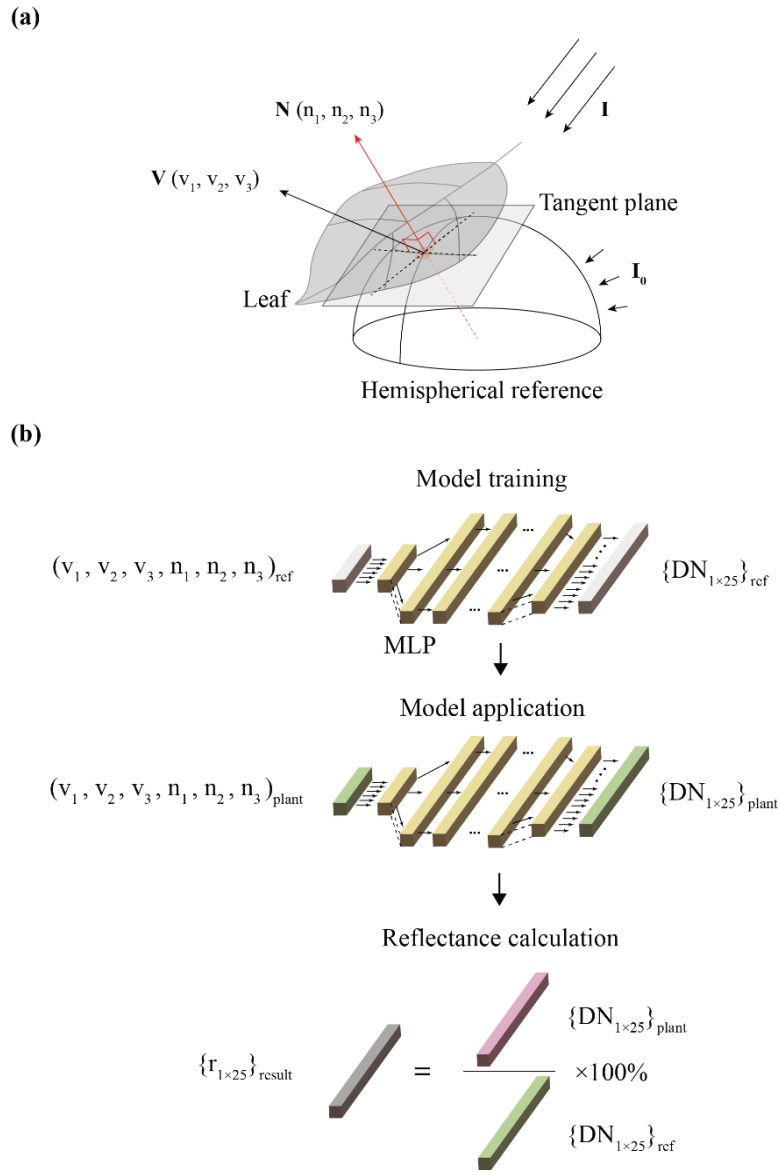

Figure 4. Schematic diagram of three-dimensional (3D) light field features and neural reference field (NeREF) training under the natural light condition. (a). The simulation of plant leaves and hemisphere tangent to each other and the relevant 3D light field features. (b). The training stage and the application stage of NeREF, and the details of the reflectance calculation method.

The augmentation of the dataset used to train the NeREF was achieved by changing the viewpoint of the reference images. We selected viewpoints at an angular interval of $15°$ on a hemisphere located at 0.5 m from the reference. The distance was set to be the same as the sight distance in plant data acquisition so that the length of all the observation direction vectors was approximately the same, which in turn helped to eliminate the observation distance factor. We also used HPSO and RRT algorithms for sorting and path planning among reachable viewpoints. Thus, the multi-sensor-equipped robotic arm can automatically complete the acquisition of the reference images.

**2.3. Fusion of MS and Depth Images**

The key steps for data fusion in the generation of 3DMPCs are to fuse MS images and depth images at each viewpoint to generate single-frame 3DMPC and to fuse multiple frames from different viewpoints to generate a complete one. The depth image and the MS image acquired from the same viewpoint were registered using the Speeded-Up Robust Features (Bay et al. 2008) and Demons (Thirion 1998) (SURF-Demons). The point clouds acquired from different viewpoints were coarsely registered using the hand-eye matrix and finely registered using the iterative closest point (ICP). The coordinate transformation of the coarse registration can be described by the following equation:

$$^{Base}P = {^{Base}_{Tcp}T} \, {^{Tcp}_{Cam}T} \, {^{Cam}P} \quad (2)$$

where $^{Base}P$ and $^{Cam}P$ are the homogeneous coordinates of the point clouds in the base coordinate system of the robotic arm and the camera coordinate system, respectively, $^{Base}_{Tcp}T$ and $^{Tcp}_{Cam}T$ are the transformation matrix between the base coordinate system and tool center point (TCP) coordinate system, TCP coordinate system and camera coordinate system, respectively. The transformation matrix $^{Base}_{Tcp}T$ was parsed from the byte stream sent by the robot arm's real-time feedback port at 125 Hz. $^{Tcp}_{Cam}T$ was calculated based on the relative position and orientation between the TCP and the RGB-D camera:

$$_{Cam}^{Tcp}T = \begin{bmatrix} 0 & 0.99955 & 0.029996 & -0.03459 \\ 0 & -0.03 & 0.99955 & 0.065924 \\ 1 & 0 & 0 & 0.12 \\ 0 & 0 & 0 & 1 \end{bmatrix} \quad (3)$$

**2.4. Experimental Evaluation**

The 3DMPCs obtained from the proposed approach were experimental tested for point cloud quality including completeness and reflectance accuracy, and chlorophyll content estimation of plant leaves. Potted perilla plants, tomato plants and rapeseed plants with different plant architectures and leaf morphological features at the seedling stage were selected to acquire 3DMPCs. Raw image data was acquired by using the UGV plant phenotyping platform, and the high-quality 3DMPCs of eighteen plants with each of six were then generated based on the proposed method. We assessed the effects of different viewpoints, fusion methods, and radiometric calibration methods on data quality. In addition, the potential of using 3DMPCs for plant phenotyping with a case study of chlorophyll content assessment was investigated.

We sampled six regions of interest (ROIs) for each plant and measured their chlorophyll content by soil and plant analyzer device (SPAD). Then we compared the absolute values of Pearson's correlation coefficients between chlorophyll content and 25-band reflectance using the flat reference-based (FR) and the hemispherical reference-based (HR) calibration, respectively. The Pearson correlation coefficient is calculated as follows:

$$\rho_{X,Y} = \frac{Cov(X,Y)}{\sigma_X \sigma_Y} \quad (4)$$

where $Cov(X,Y)$ is the covariance between X and Y, $\sigma_X$ and $\sigma_Y$ are the standard deviations of X and Y. In addition, we established a SPAD prediction model using the partial least squares regression (PLSR) and compared the prediction results after FR and HR calibration. The coefficient of determination ($R^2$) and the average root mean square error (RMSE) were selected as the evaluation metrics.

**2.4.1. Point Cloud Completeness**

The completeness of the point cloud acquired from FPVs, VVPs and AVPs was evaluated. The coverage was defined as the ratio of the space occupied by incomplete and complete point clouds as follows:

$$\text{Coverage} = \frac{V_{fusion}}{V_{gt}} \tag{5}$$

where $V_{fusion}$ is the number of voxels of the fused point cloud and $V_{gt}$ is the number of voxels of the ground truth obtained from the laser scanner. Both the fused point clouds and the ground truth were down-sampled using the voxel-grid filter to ensure the same spatial occupancy of a single point.

### 2.4.2. Reflectance Accuracy

We sampled 6 ROIs for each plant and measured their reflectance as the ground truth using an ASD spectrometer (Analytical Spectral Device, Inc., Boulder, CO, USA) equipped with the leaf clip. The RMSE and the Euclidean distance (ED, a classic spectral similarity metric) range of the spectra at the ROI were selected as the evaluation metrics to measure the spectral accuracy and the multi-view measurement stability respectively:

$$\text{RMSE} = \frac{1}{N}\sum_{v=1}^{N}\sqrt{\frac{1}{M}\sum_{b=1}^{M}\left(s_v^b - s_{gt}^b\right)^2} \tag{6}$$

$$\text{ED range} = max\left(\sqrt{\frac{1}{M}\sum_{b=1}^{M}\left(s_m^b - s_n^b\right)^2}\right) \tag{7}$$

where $s_v^b$ is the reflectance value of viewpoint $v$ at band $b$, $s_{gt}^b$ is the reflectance value measured by ASD at band $b$, $M$ is the number of bands, $N$ is the number of viewpoints, and $m, n \in [1, N]$.

### 2.4.3. Configurations for Generating 3DMPCs

The NBV planning program was written in C++ in the environment of Qt 5.2.0. The application programming interfaces (APIs) of the point cloud processing and path planning were Point Cloud Library (PCL) 1.8.0. and MoveIt in the Melodic version of the Robot Operating System (ROS). The Training of NeREF was performed with Deep Learning Toolbox (v14.0) in MATLAB 2020a software (MathWorks, Natick, Massachusetts, USA) on the same computer as previously mentioned. The input and output layer sizes were determined by the number of 3D light field features and the number of spectral bands, respectively. The number of middle layers and the number of neurons in each middle layer were set to 8 and 128 according to the empirical formula and the preliminary experiment. We obtained a training set with 394390 samples, a validation set with 84535 samples, and a test set with 84535 samples after automated data collection and classification. The dataset was divided into the 70% of training, 15% of validation, and 15% of testing sets, and the mean square error (MSE) was also calculated. Due to the scale heterogeneity of the data, the min-max normalization of the data was performed before model training. The configuration of the computer consisted of Intel(R) Core (TM) i7-9850H @ 2.60 GHz, a 64-bit operating system, and 64GB of ECC RAM. The training took 5 h 20 min, went through 338 epochs.

3. Results

3.1. Performance of NBV Planning

3.1.1. Plant-oriented NBV

Figure 5 illustrates the NBV estimation results for each step. Both AABB and OBB abstracted the target as a cuboid that provided many faces for generating suitable camera poses. These poses ensured that the camera was perpendicular to one of these faces for the maximum current information gain. Each face of the AABB had a corresponding viewpoint (i.e., FVP) (Figure 5a, 5d, and 5g) to acquire the sufficient structural and textural information from side and top views. In contrast, the thickness of the leaves was much smaller than their length and width, and the front of the OBB contained much more information than the sides. Therefore, only a frontal viewpoint of the OBB (i.e. VVP) was required to obtain complete information about the leaf (see Figure 5b, 5e and 5h). The above-mentioned viewpoints designed for the whole plants or leaves did not consider their relative positions and topological relationships. The convex hull bridged the gap between the outer and the inner layer of the plant, as the larger polygons of its surface acted as the windows for observing the internal organs. Figure 5c, 5f, and 5i show the estimated AVPs based on the VVPs. Table 1 shows the efficiency of the adaptive algorithm in terms of the inference and execution time. The inference time for the AVP estimation was the longest among the three steps due to more iterations, nearly 25 times more than that for the FVP estimation. Nevertheless, the time was still acceptable in practical use as it took only a few seconds. The time spent at each step of the viewpoint sorting and path planning did not exceed 0.4 s. Additionally, the execution time for all steps was almost less than 26 s at the joint speed of 1.55 rad/s. The total acquisition time (sum of inference time and execution time) of the proposed method for a single plant was approximately 1 min, reduced by 11.6 s averagely compared to the non-HPSO method. Overall, the inference time of the proposed adaptive algorithm is appropriate for the application, and the execution is more efficient after implementing HPSO sorting.

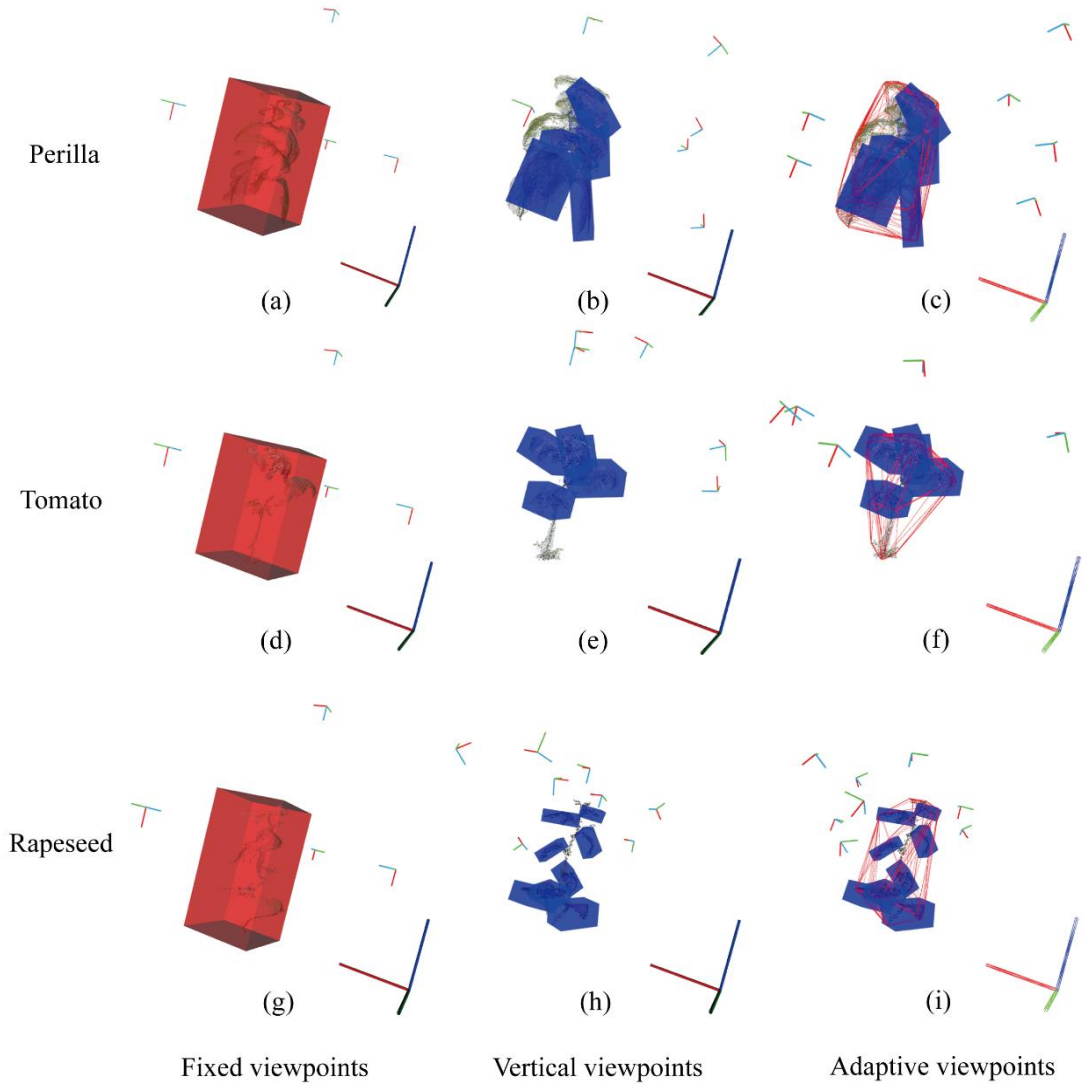

Figure 5. Illustration of the next-best-view (NBV) estimation results for each step. The pose in the lower right corner of each sub-figure is the base coordinate system of the robotic arm, and the other poses are the predicted coordinate systems of the depth camera for each viewpoint. The red and blue box represents the axis-aligned bounding box (AABB) and oriented bounding box (OBB), respectively.

Table 1. Inference and execution time of each step for the proposed adaptive algorithm.

| Plants | Perilla | Tomato | Rapeseed |
| --- | --- | --- | --- |

|  |  |  |  |  |  |
|---|---|---|---|---|---|
| Inference time/s | NBV estimation | FVPs | 0.060 | 0.058 | 0.061 |
|  |  | VVPs | 0.154 | 0.135 | 0.182 |
|  |  | AVPs | 1.511 | 1.167 | 1.986 |
|  | Path planning | FVPs | 0.248 | 0.225 | 0.244 |
|  |  | VVPs | 0.290 | 0.242 | 0.360 |
|  |  | AVPs | 0.283 | 0.253 | 0.348 |
| Execution time/s |  | FVPs | 17.1 | 16.2 | 17.5 |
|  |  | VVPs | 21.2 | 19.4 | 25.7 |
|  |  | AVPs | 20.4 | 18.0 | 24.9 |
| Total time/s |  |  | 61.3 | 55.7 | 71.3 |

Note: The joint speed at the execution was 1.55 rad/s. NBV next-best-view, FVP fixed viewpoint, VVP vertical viewpoint, AVP adaptive viewpoint.

### 3.1.2. Reference-oriented NBV

Figure 6a illustrates the viewpoints for reference image acquisition. There were 79 reachable viewpoints rendered in RGB, and the unreachable viewpoints, due to the limitation of the movement space of the robotic arm, were rendered in white. Figure 6b presents the fitness curve of HPSO applied to sort the viewpoint. The fitness decreased from 191 rad to 100 rad in 93 s after 5000 iterations. We adopted the results after 5000 iterations to generate the usable well-ordered viewpoints. When applying RRT for path planning, the execution times of the robotic arm based on the sorted and unsorted viewpoints were 184 s and 215 s, respectively. Overall, the use of HPSO and RRT allowed the robotic arm to complete the reference image acquisition for all viewpoints in less than 5 minutes.

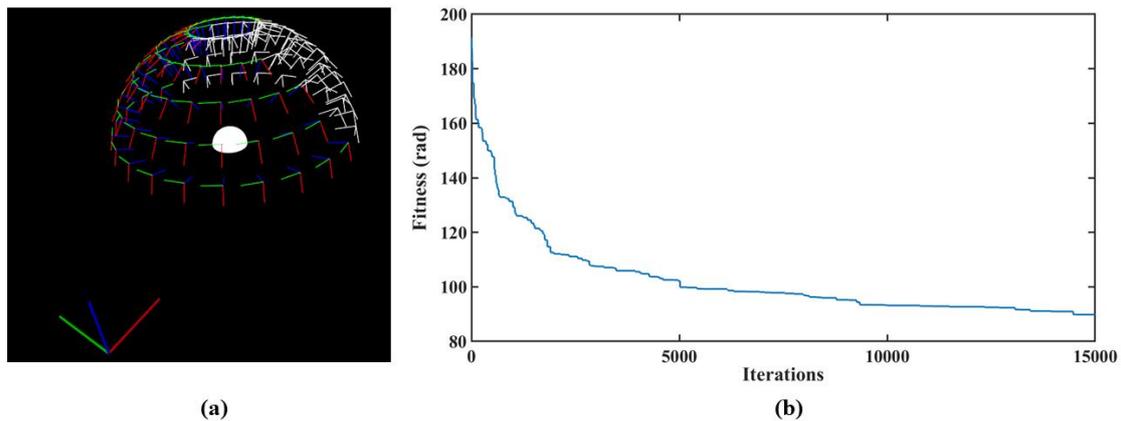

(a)　　　　　　　　　　(b)

Figure 6. Viewpoints for acquiring reference images, and their sorting process. (a) is the visualization of the viable viewpoints. Unreachable viewpoints are shown in white.

(b) is the fitness curve of the hybrid particle swarm optimization (HPSO) applied to sort the viable viewpoints.

### 3.2. Point Cloud Quality Assessment

Figure 7 shows the visualization of the completeness of the leaf point clouds from different viewpoints. The occlusion and limited FOV led to the loss of point clouds and spectral information in the best FVP ($FVP_{best}$) and VVP, which could cause variation in the multi-sensor image registration. As shown in Figure 8, the coverage of point clouds with both RGB and MS information was lower than that of point clouds with only RGB information due to the difference in FOV. Despite this, they both benefit from the combinations of viewpoints. For the single type of viewpoints, all coverage rates of RGB and RGB+MS point clouds are below 0.85 and 0.70, respectively. While the coverage rates of points significantly increased by at least 9.7% and 7.4% with different combinations of viewpoints. Specifically, the coverage of the RGB and RGB+MS point clouds acquired from the combination of FVPs, VVPs, and AVPs reached the highest with the rate of 0.99 and 0.82, respectively, and the combination of the FVPs and VVPs achieved the lowest coverage rate. The two highest values are on average 22.1% and 25.0% higher than those acquired from FVPs only. In summary, the multi-sensor data quality can be significantly improved with the complementarity of the point cloud completeness at different viewpoints.

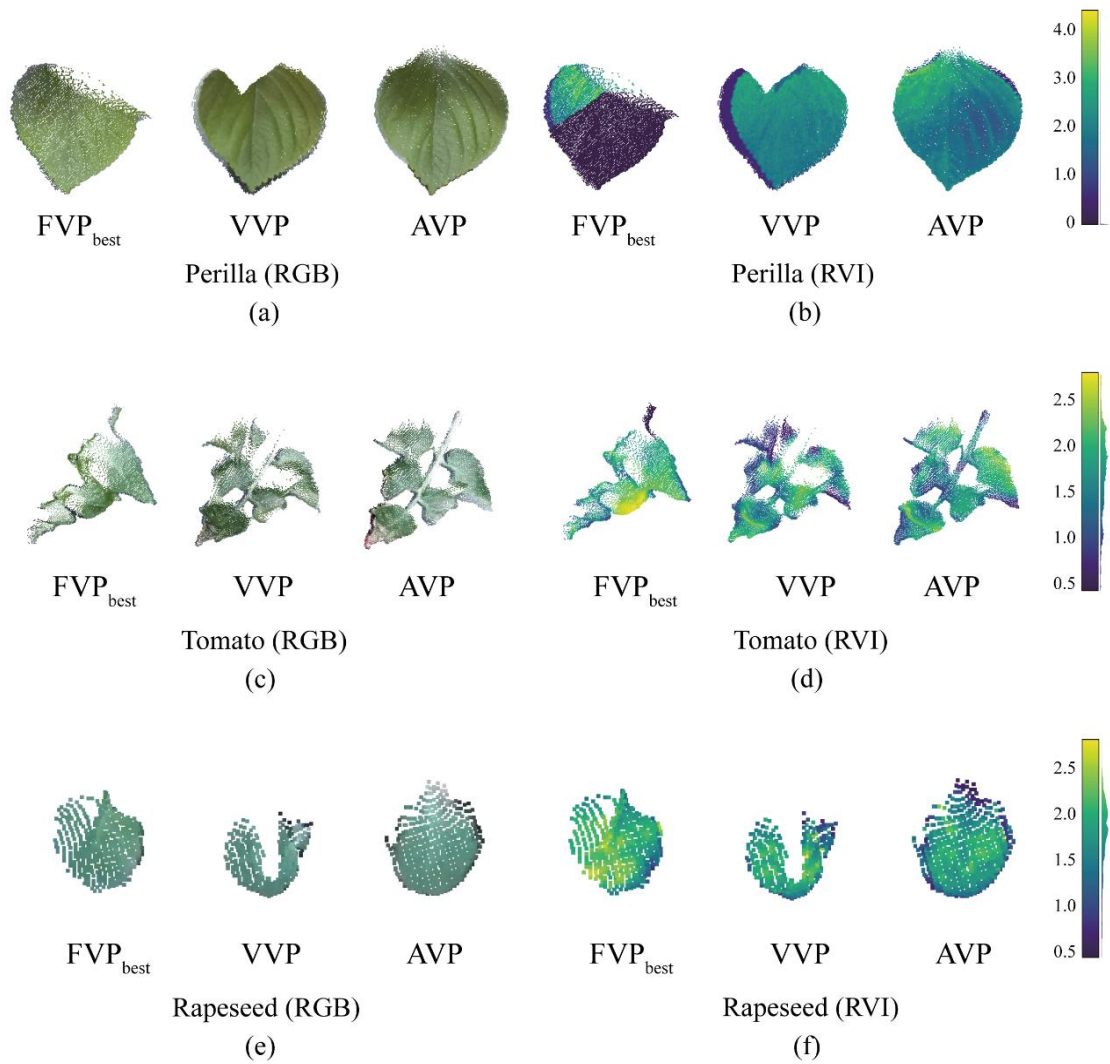

Figure 7. Visualization of the completeness of the (a, c, e) RGB and (b, d, f) ratio vegetation index (RVI, ratio of reflectance between 753.8 nm and 666.8 nm) point clouds of perilla, tomato and rapeseed for the best fixed viewpoints (FVP$_{best}$), the vertical viewpoints (VVP), and the adaptive viewpoints (AVP).

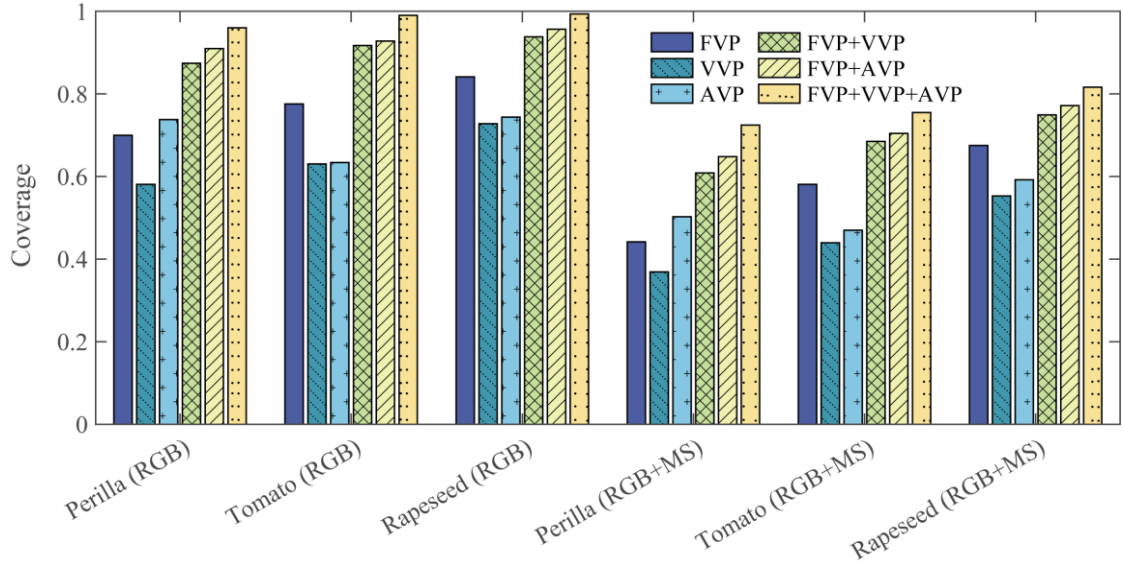

Figure 8. Coverage of the fused point clouds of the whole plant from the fixed viewpoints (FVP), vertical viewpoints (VVP), adaptive viewpoints (AVP), and their combinations. The data include RGB point clouds (only RGB information) and RGB+MS point clouds (both RGB and multispectral information).

### 3.3. Evaluation of Radiometric Calibration

The training loss, error distribution, and results of spectral DN value prediction of NeREF are presented in Figure S1. The NeREF was well-trained and performed well on the test set with $R^2$ of 0.982 and RMSE of 0.051 (Figure S1e), indicating that the NeREF can achieve an accurate prediction of DN values of the reference, which is crucial for radiometric calibration. Figure 9 shows the comparison of RMSE and the ED range of reflectance spectra using FR and HR calibration at all ROIs. The average RMSEs for the HR-calibrated reflectance spectra of perilla, tomato and rapeseed were 0.08, 0.07 and 0.10, respectively, which were significantly lower than the FR-calibrated ones with the RMSE values of 0.22, 0.19 and 0.20, respectively. The average ED range of the HR-calibrated reflectance spectra was 0.044, which is higher than the FR-calibrated one of 0.029. The above results indicate that NeREF significantly improved the spectral accuracy although it amplified the differences in spectral data of the same ROI collected from different viewpoints. Since neither the plant nor the hemispherical reference was actually a perfect Lambertian and NeREF itself had prediction errors, there would be differences in the DN values collected from different viewpoints of the same ROI. Therefore, considering that the ED range measures the most extreme case, an average of 0.044 is actually acceptable. Examples of the reflectance of a single-frame plant 3DMPC at the wavelength of 740.7 nm after FR and HR calibration were presented in Figure 10. By using FR calibration, a large reflectance distortion caused by the heterogeneity of plant-light interactions observed at different viewpoints still exists in perilla, tomato and rapeseed plants, and the variation in one single leaf reached 0.33. However, the HR-calibrated reflectance was relatively uniform within the leaf ranging between 0.5 and 0.7, which is consistent with the ASD measurements in multiple regions of the same leaf (the maximum variation is 0.07).

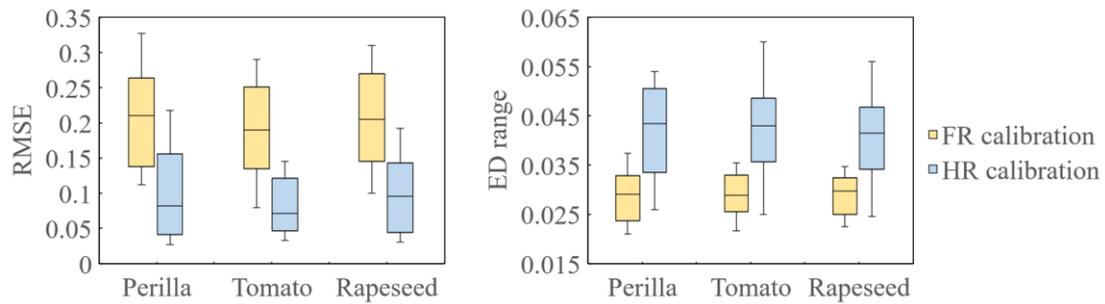

Figure 9. Comparison of root mean square error (RMSE) and Euclidean distance (ED) range of reflectance spectra using the flat reference-based (FR) calibration and the hemispherical reference-based (HR) calibration at all regions of interest (ROIs).

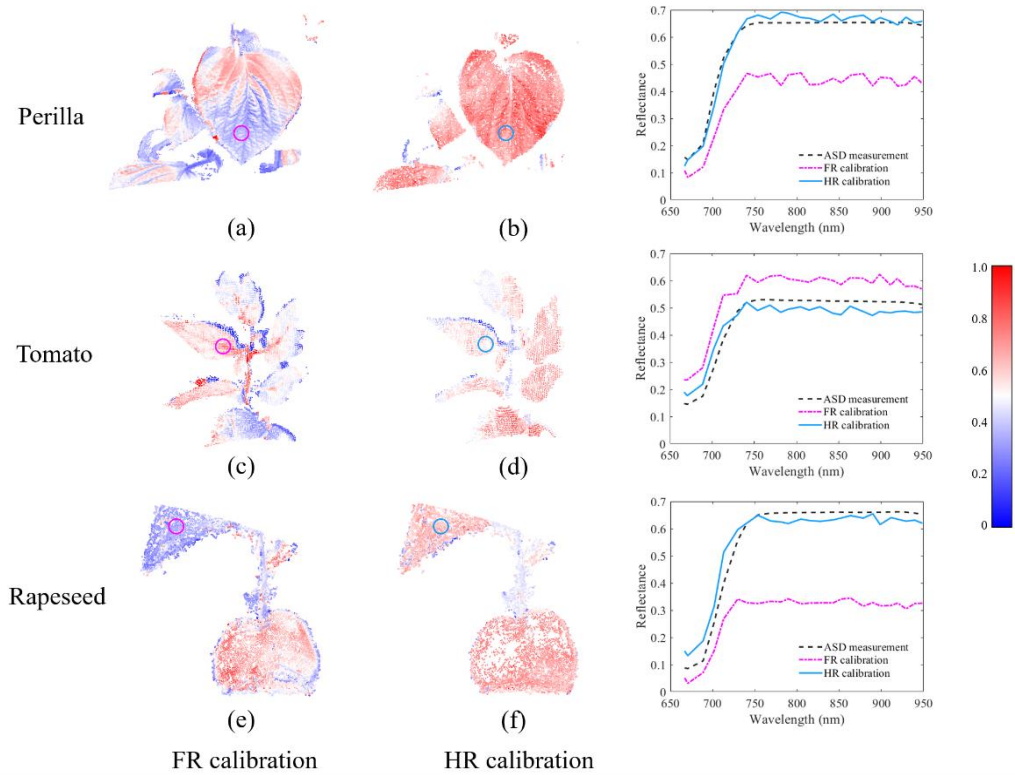

Figure 10. Visualization of the radiometric calibration results for a single-frame three-dimensional multispectral point cloud (3DMPC) of perilla, tomato and rapeseed at 740.7 nm, and the reflectance spectra at all regions of interest (ROIs) obtained using different calibration methods. (a), (c) and (e) are the results after the flat reference-based (FR) calibration, and (b), (d) and (f) are the results after the hemispherical reference-based (HR) calibration.

### 3.4. Applications in Chlorophyll Content Estimation

Figure 11 presents the prediction results of plant SPAD value by using PLSR with multispectral point clouds. It was found that the performance of PLSR model with the HR-calibrated data was much better than that obtained with the FR calibration. The $R^2$ increased from 0.82 to 0.88 with the RMSE decreased by 18.27% for predicting SPAD value of perilla leaves. In the case of tomato, the $R^2$ increased from 0.84 to 0.91, and the RMSE decreased by 25.58%. For rapeseed, the $R^2$ rose from 0.81 to 0.89 with a corresponding 19.89% decrease in RMSE. The results demonstrated that the proposed approach successfully improved the quality of multispectral point clouds and achieved a good performance to evaluate the biochemical traits in the 3D spatial domain no matter how complex the plant structure (the correlation between chlorophyll content and reflectance in each band obtained using different calibration methods is shown in Figure S2). More interestingly, the spatial distribution of the SPAD values in the whole plant can be generated by using the high-quality 3DMPCs as shown in Figure 12. The yellow or decayed leaves were easily identified with the low SPAD values, which was consistent with their RGB images.

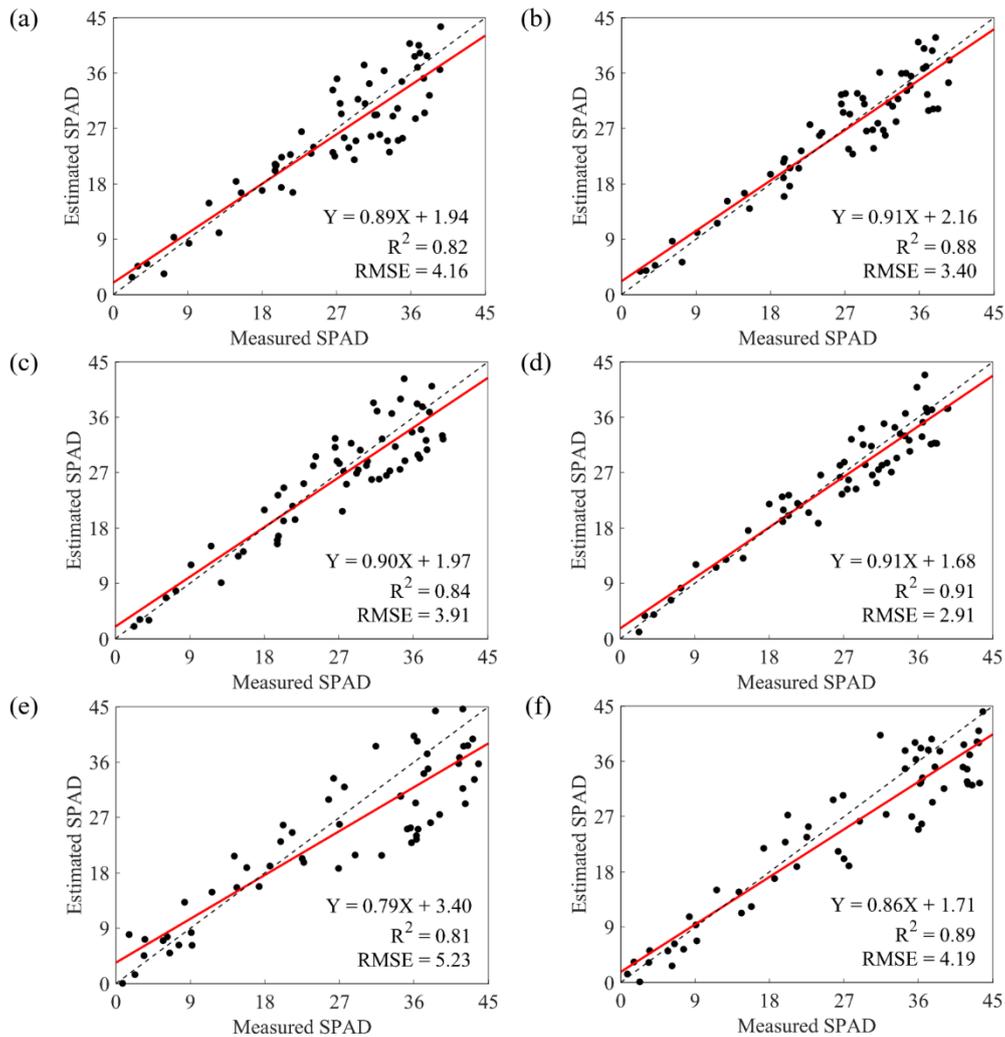

Figure 11. Prediction results of plant SPAD using partial least squares regression (PLSR). (a) and (b) are perilla SPAD prediction using the flat reference-based (FR) and the hemispherical reference-based (HR) calibration. (c) and (d) are tomato SPAD prediction using FR and HR calibration. (e) and (f) are rapeseed SPAD prediction using FR and HR calibration.

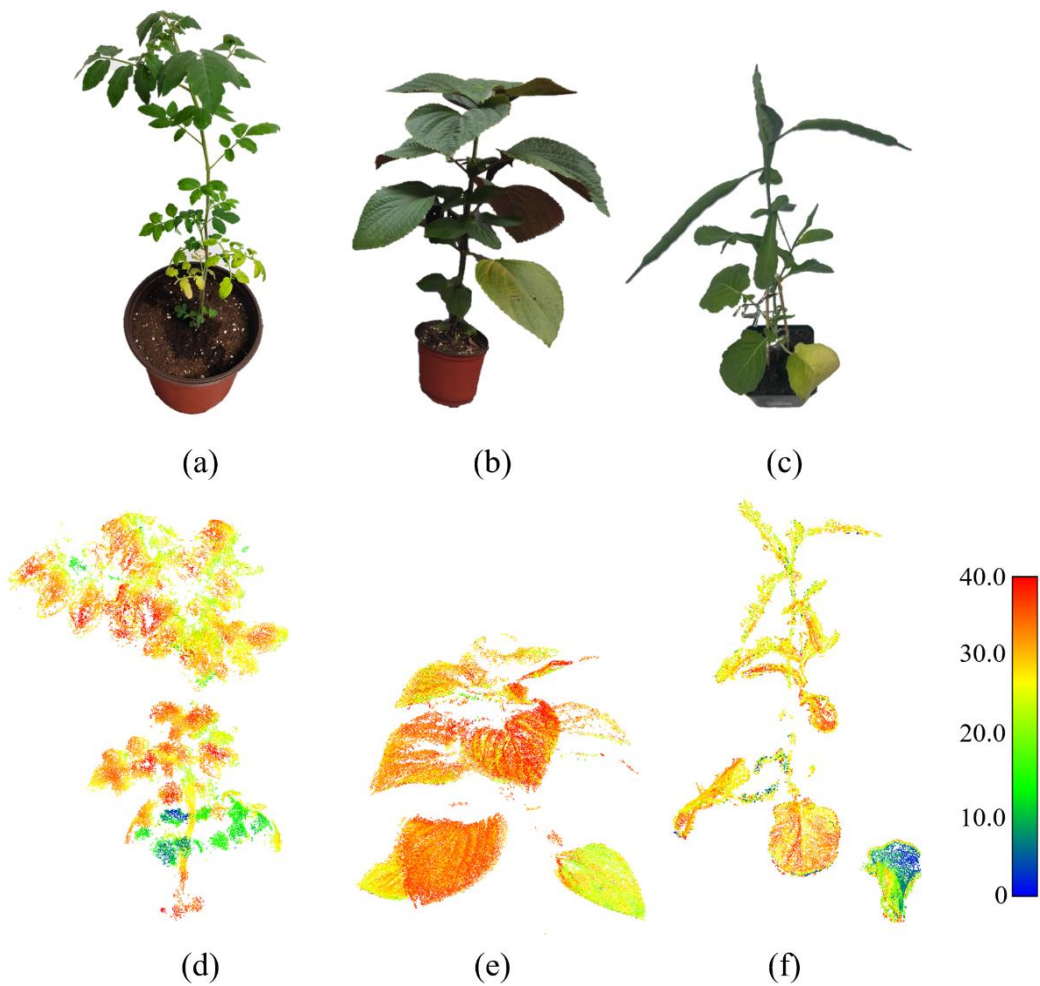

Figure 12. (a, b, and c) RGB images and (d, e, and f) visualization of prediction results of SPAD distributions along the whole plants (i.e., tomato, perilla and rapeseed plants) by using partial least squares regression (PLSR) with the three-dimensional multispectral point clouds (3DMPCs).

## 4. Discussion

### 4.1. Improvement of Target Integrity with Better Main Direction Estimation

The main direction in NBV estimation that is generally perpendicular to the projection plane contains the most information about the target, and the unobstructed viewpoint of the target can then be estimated by constructing a unit direction hemisphere. While different leaf morphologies could influence the estimation of the main direction, and eventually lead to the variation in the coverage of the fused point clouds. As shown in Figure 8, the difference in the coverage of the fused point clouds of perilla from VVPs and AVPs was significantly greater than that of tomato and rapeseed plants. In this study, we used PCA to estimate the main directions of the targets, i.e., the normal directions to the largest faces of the OBBs. Perilla leaves are wide and less curled, and their OBB length and width are much smaller than the OBB height. The main directions can then be clearly determined to be perpendicular to the leaf surfaces, and the potential occlusions can be effectively avoided when implanting AVPs. However, tomato leaves are irregularly shaped pinnate compound leaves with leaflets that are ovate or elliptic, generally borne in pairs from the base of the main petiole toward the apex, while the apical leaflets are borne on a short petiole (Tattersall et al. 2005). The petiole acts as a conduit for nutrients and water between the leaf and the stem and allows leaves to change the direction, making the tomato plants more susceptible to deformation. Consequently, it could cause the OBB length and width of tomato leaves to not always be greater than their OBB height (the OBBs of perilla and tomato leaves have obvious shape differences in Figure 5), leading to uncertainty in the estimation of the main directions. Hence, when tomato leaves are not heavily occluded in the main directions, AVPs may contribute little to improving the coverage rate of point clouds. As for rapeseed plants, there are obvious differences in the vertical distribution of the canopy (the OBB differences between the upper and lower leaves of the rapeseed plant are large in Figure 5h and 5i). The upper leaves are narrow and flexible, and the lower leaves are wide and flat, similar to tomato leaves and perilla leaves respectively, but with much less occlusion. Therefore, the coverage rates of the fused point clouds from FVPs, VVPs and AVPs are higher than those of the other two plants (see Figure 8).

The uncertainty of the main directions caused by the area proximity of each OBB face could be resolved by adding multiple main directions for a single OBB based on the area ordering and the orientation, or even the customization. Although this would increase the number of potential NBVs, it can effectively improve the target integrity in adaptive data acquisition and extend the proposed adaptive algorithms to different plant organs, such as stems, fruits, and other easily shaded organs. In addition, some sensing tasks that require denser viewpoints can apply the proposed NBV estimation method to obtain richer information, e.g., using a laser scanner to obtain dense 3D point clouds of plants, which requires more intersections between the laser and the plant surface (Chaudhury et al. 2019; Chaudhury et al. 2015; Paulus 2019). This can be achieved by setting a scan path that includes all candidate NBVs. Overall, the number of the candidate NBVs can be increased by adding main directions to expand the application scenarios of the adaptive algorithm in different sensing targets and sensing tasks.

**4.2. Pre-planning for Faster Inference**

The number of NBVs has a significant effect on the inference time, especially for reference-oriented NBVs. The sorting and path planning of dozens of viewpoints took nearly two minutes throughout the process, while the longer elapsed time may not be conducive to maintaining the stability of the light field. For plant-oriented NBVs, although the sorting and path planning time was not long due to the small number of NBVs, the addition of the main directions for single OBB (mentioned in section 4.1) may also lead to the excessive inference time in the scenarios where the sizes of each facet of the target OBB are close to each other. Development of a database with pre-sample viewpoints of the robotic arm and the trajectory planning could be a solution (Wu et al. 2019). The estimated NBVs can be matched with the pre-sampled viewpoints to reduce the computational complexity, and the planned trajectories can then be queried in real-time during image acquisition.

**4.3. Importance of Light Field Representation**

The light field is a common representation for describing the positional and orientational relationships between the light source, plant, and optical sensor (Sitzmann et al. 2021). Despite the complexity of the real light conditions, it can be simulated with the combination of multiple ideal light fields, such as point-source, parallel, and collocated light fields (Figure 13). The point-source light field is the most fundamental form of light field, and any light field can be viewed as a composition of point-source light fields. However, the representation of a point-source light field is relatively intricate, necessitating prior knowledge of the light source's location to determine the light field features, and multiple hemispherical references are also required to augment the data. This limitation confines its applicability to indoor phenotyping platforms with controlled lighting and imaging conditions as reported in our previous study (Xie et al. 2023). Additionally, the representation of a point-source light field constrains the arbitrary selection of viewpoints, as changes in the relative positions of the light source and camera significantly alter the incident and observation vectors. Consequently, NeREF trained on data acquired from specific viewpoints cannot be effectively applied to predict DN values at new viewpoints unless there exists symmetry among viewpoints and light sources (such as a circular distribution of viewpoints or rotating plants only), thus eliminating the differences in the relative positions. In contrast, a parallel light field is relatively simple to be implemented at various scenarios. The uniformity of incident light and adaptive data acquisition enable the training of NeREF using multi-view data with just a single hemispherical reference. This makes it more suitable for outdoor scenes, particularly when deployed on the mobile phenotyping robots in the greenhouse or open-field.

It was also noticed that the shadow regions without a good lightening condition exist in both point-source and parallel light fields, which could reduce the estimation accuracy of DN value of the reference. One possible solution to solve this issue is to mount both the light source and the camera on the robot arm (Figure 13c). The collocated setting results in the same direction vector of incident light and observation ($\boldsymbol{I} = \boldsymbol{V}$), undoubtedly simplifying the light field representation and eliminating the shadows from each viewpoint (Bi et al. 2020). With a fixed incidence and observation distance, the light conditions are almost the same for each viewpoint, so it is possible to perform data augmentation at only one viewpoint to obtain a good estimation of DN values of hemispherical references.

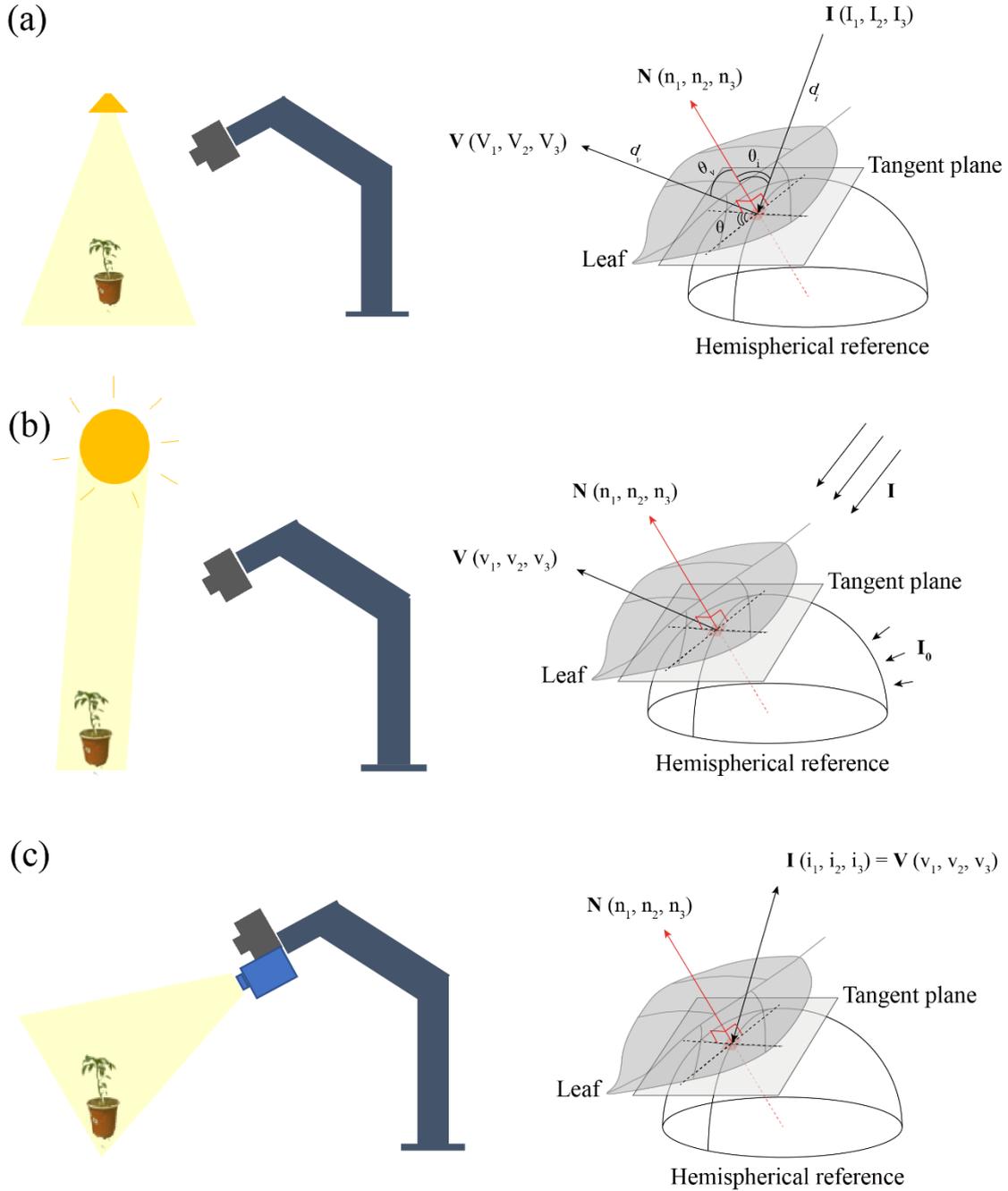

Figure 13. Configurations of three-dimensional multispectral point clouds acquisition at (a) the point-source light field, (b) parallel light field, and (c) collocated light field. *I*, *V*, and *N* are the direction vector of the incident light, the observation direction vector, and the normal vector of the surface, respectively. $I_0$ is the diffuse light. For the specific value of each dimension, lowercase means that the distance value does not need to be considered, and the vector length is 1.

**4.4. Error Analysis in Radiometric Calibration with Advanced Neural Fields**

The proposed NeREF for per-ray DN value prediction is a novel implicit neural scene representation that directly parameterizes the light field of the hemispherical reference via a neural network. The low RMSE and reasonable ED range for the calibrated reflectance spectra of perilla, tomato and rapeseed has shown the superior performance of HR calibration with advanced neural fields compared to the common FR calibration. However, the optical properties of the measured target and the accuracy of representation may affect the prediction accuracy of DN values. It is assumed that the measured target is approximated to an opaque and hard surface object, since each ray has only one intersection point with the surface to encode a single surface property in NeREF (Zhang et al. 2021b). For the plant materials, light can be both reflected from or transmitted into the plant leaves, and the transmittance component could add a systematic error to the radiometric calibration (the HR-calibrated spectral RMSE of perilla, tomato and rapeseed in Figure 9 is still not 0). The self-occlusion of plant organs would be another challenge to model rays from the light field, as ray is assumed not to pass through the object surface in the light field representation (Li et al. 2021). There are small parts of perilla point clouds that were deleted due to extreme anomalies in reflectance after HR calibration in Figure 10b and 12e, most of which are located in occluded regions. Besides, 3D light field feature extraction relies on accurate surface normal estimation, while edge effects caused by depth value anomalies, leaf edge curling, etc. will reduce the already low accuracy of leaf edge normal estimation, leading to feature mismatches and low calibration accuracy (see partial reflectance and chlorophyll content anomalies at the leaf edges shown in Figure 10 and 12).

Furthermore, there is an assumption that the hemispherical reference and the plant have the same bidirectional reflectance distribution function (BRDF) when estimating the DN value of hemispherical reference under the same light field, which could be violated due to the specular reflectance caused by the waxy layers on the leaf surface (Holmes and Keiller 2002). Likewise, the reference is not an ideal Lambert radiator. These can lead to abnormal DN value predictions as shown in Figure S1c-S1e. The bidirectional scattering distribution function (BSDF), containing both non-Lambertian BRDF and bidirectional transmittance distribution function (BTDF) could be another option for future studies.

In addition, the challenge of plant radiometric calibration for more complex light conditions must be taken into consideration, since in most cases it is not easy to manually generalize light field characteristics (e.g., the position, orientation, size, etc. of the light source). We can draw inspiration from the NeRF-based relighting and scene material editing, whose goal is to estimate the material properties and the illumination of a surface or participating media from sparse measurements such as images. These applications leveraged the neural fields to regress the required rendering properties such as volume density, normal, visibility, albedo, and reflectance properties (e.g., parameters of the manually selected analytic BRDF based on prior knowledge of the lighting, or the reflectance functions learned from real-world BRDFs), from the 3D locations in the entire volume for differentiable ray marching (Srinivasan et al. 2021; Zhang et al. 2021a; Zhang et al. 2021b). If used for plants, these techniques will facilitate end-to-end accurate estimation of properties including reflectance and 3D geometry of plants.

## 5. Conclusions

This study shows that it is promising to apply adaptive data acquisition and NeREF to generate high-quality 3DMPCs of plants under natural light conditions. We proposed an efficient pipeline for NBV planning, i.e., to estimate NBVs, and plan paths among these viewpoints based on HPSO and RRT. We also designed a novel implicit neural scene representation named NeREF that directly parameterizes the light field of the hemispherical reference via a neural network, and facilitates the efficient acquisition of spectral DN of the reference for the plant in the same light field. The results show that NBV planning is competitive and that adaptive data acquisition can be completed within an acceptable time frame, which improves data integrity and reduces subjectivity. NeREF achieves a good prediction of the spectral DN of the hemispherical reference. It calibrates the reflectance spectra of plants well and effectively eliminates the illumination effects in the natural light condition. The accuracy of plant chlorophyll content estimation can be improved by using high-quality 3DMPCs generated by the above method. Overall, our method enables the generation of high-quality plant 3DMPCs for efficient phenotypic measurements under natural light conditions.

Continuous measurements are viable without notable changes in light conditions over a short period of time. Further research will be conducted on aspects such as NBV planning for different plant organs, and end-to-end plant radiometric calibration under more complex light conditions based on neural fields.

**Acknowledgments**

**Supplementary Materials**

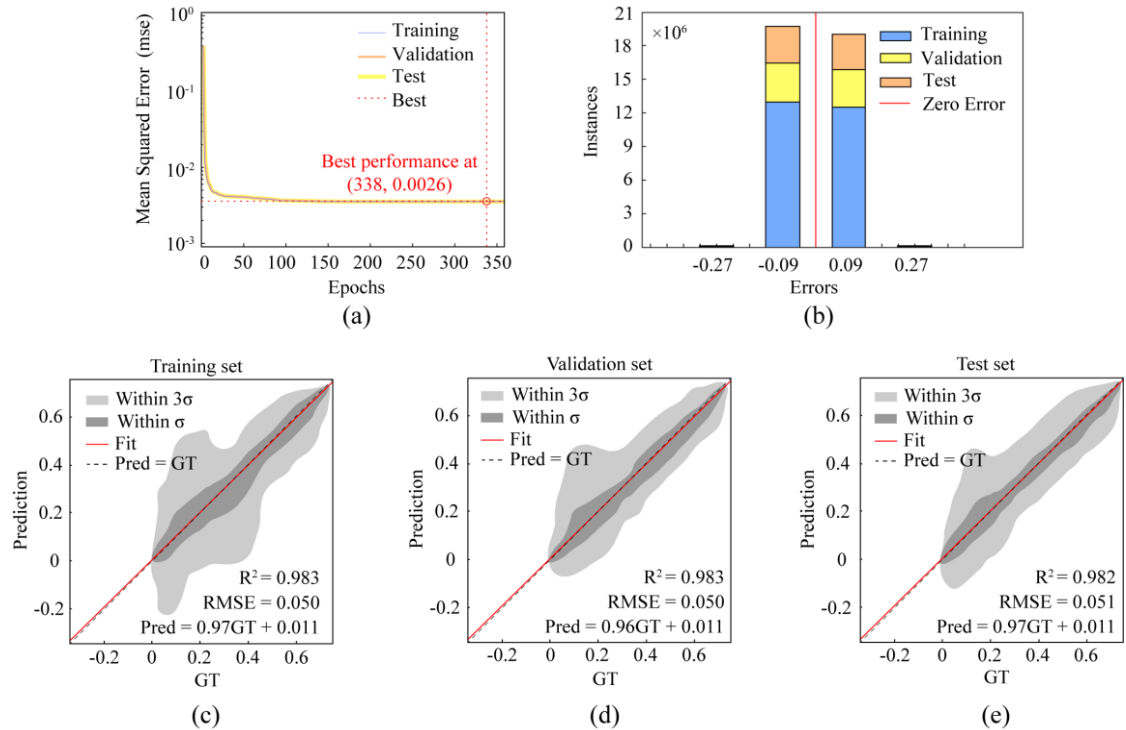

Figure S1. Performance of neural reference field (NeREF) for spectral digital number (DN) values prediction. (a). The curve of the loss function changing with the training epoch. (b). The error distribution histogram of the DN values predicted by NeREF. (c-e). Regression results between predictions and ground truth in training set, validation set and test set.

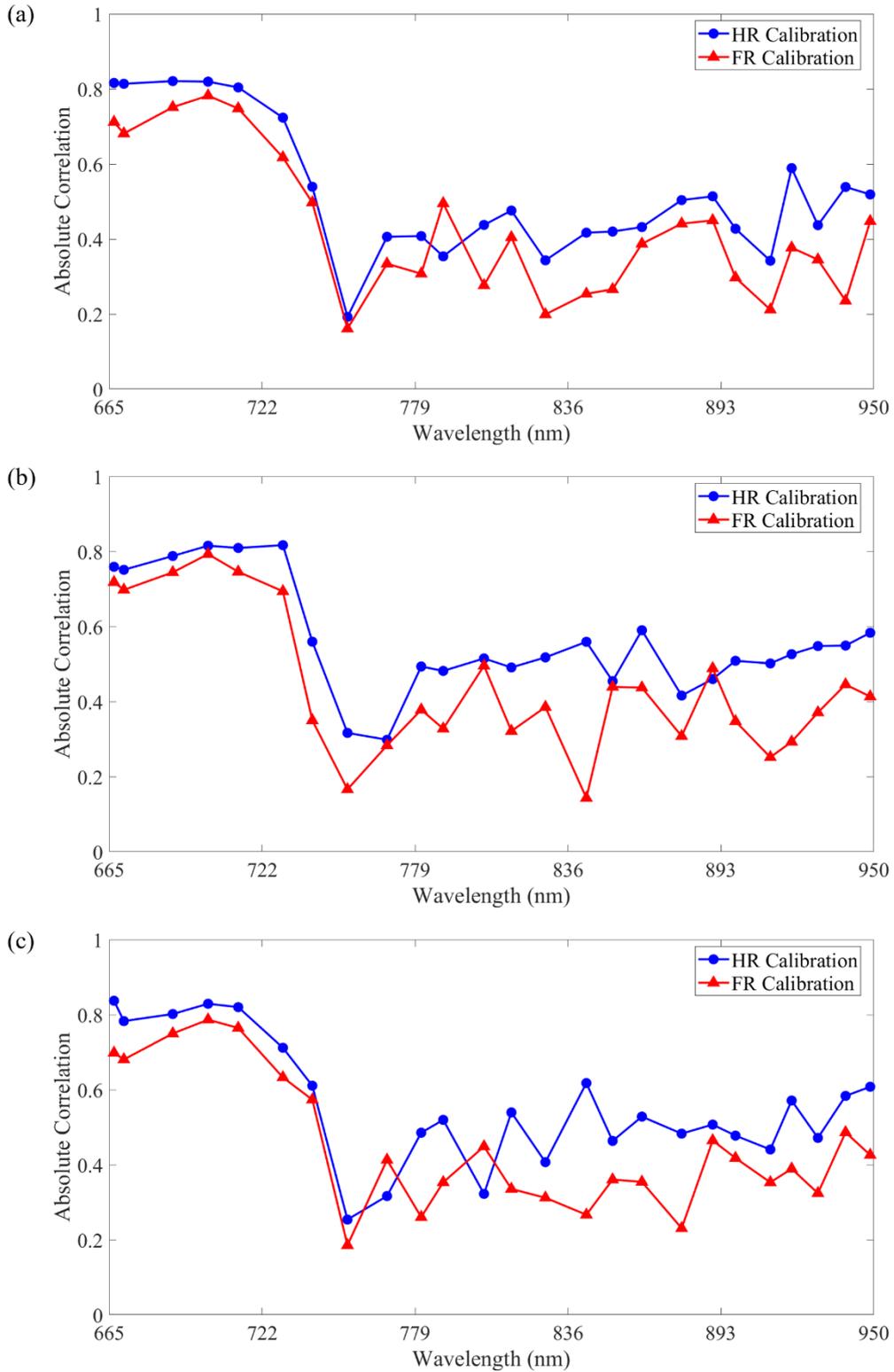

Figure S2. Absolute values of Pearson's correlation coefficients between chlorophyll content and 25-band reflectance using the flat reference-based (FR) calibration and the hemispherical reference-based (HR) calibration for tomato (a), perilla (b) and rapeseed (c).